\documentclass[runningheads]{llncs}
\usepackage{cite}
\usepackage{amsmath,amssymb,amsfonts}
\usepackage{algorithmic}
\usepackage{graphicx}
\usepackage{textcomp}
\usepackage{xcolor}
\usepackage{lipsum}
\usepackage{caption}
\usepackage{subcaption}
\def\BibTeX{{\rm B\kern-.05em{\sc i\kern-.025em b}\kern-.08em
    T\kern-.1667em\lower.7ex\hbox{E}\kern-.125emX}}
\usepackage{placeins}
\usepackage{tabularx}
\usepackage{longtable}
\usepackage{fontawesome}
\usepackage{soul}
\usepackage{tcolorbox}
\usepackage{colortbl}
\usepackage{changepage}
\usepackage{framed,enumitem}
\usepackage{setspace}
\usepackage{tcolorbox}
\usepackage{wrapfig}
\usepackage{hyperref}
\newcommand{\interviewquote}[2]{
 \def\FrameCommand{%
    \hspace{0pt}%
    {\color{blue}\vrule width 1.5pt}%
    {\color{white}\vrule width 4pt}%
    \colorbox{white}
  }%
  \MakeFramed{\advance\hsize-\width\FrameRestore}%
  \noindent\hspace{-4.55pt}
  \begin{adjustwidth}{}{7pt}
  {\footnotesize ``\emph{#1}'' - {#2}}\vspace{0.5pt}\end{adjustwidth}\endMakeFramed%
}
\begin{document}
\title{Requirements Engineering for Automotive Perception Systems: an Interview Study}
\titlerunning{RE for Automotive Perception Systems}

\vspace{-0.3cm}
%
\author{Khan Mohammad Habibullah\inst{1} \and Hans-Martin Heyn\inst{1} \and Gregory Gay\inst{1} \and Jennifer Horkoff\inst{1} \and  Eric Knauss\inst{1} \and Markus Borg\inst{2} \and Alessia Knauss\inst{3} \and Håkan Sivencrona \and Polly Jing Li\inst{4}   }
\authorrunning{Habibullah et al.}
%
\institute{Chalmers $\mid$ University of Gothenburg, Sweden \and
Lund University, Sweden \and
Zenseact AB, Sweden \and Kognic AB, Sweden\\
\email{\{khan.mohammad.habibullah, hans-martin.heyn, jennifer.horkoff, eric.knauss\}@gu.se, greg@greggay.com, markus.borg@cs.lth.se, \{alessia.knauss, hakan.sivencrona\}@zenseact.com, polly.jing.li@kognic.com}}
\maketitle              

\vspace{-0.3cm}

 \begin{abstract}

\textbf{Background:} Driving automation systems (DAS), including autonomous driving and advanced driver assistance, are an important safety-critical domain. DAS often incorporate perceptions systems that use machine learning (ML) to analyze the vehicle environment. 
\textbf{Aims:} We explore new or differing requirements engineering (RE) topics and challenges that practitioners experience in this domain. 
\textbf{Method:} We have conducted an interview study with 19 participants across five companies and performed thematic analysis. 
\textbf{Results:} Practitioners have difficulty specifying upfront requirements, and often rely on scenarios and operational design domains (ODDs) as RE artifacts. Challenges relate to ODD detection and ODD exit detection, realistic scenarios, edge case specification, breaking down requirements, traceability, creating specifications for data and annotations, and quantifying quality requirements. 
\textbf{Conclusions:} Our findings contribute to understanding how RE is practiced for DAS perception systems and the collected challenges can drive future research for DAS and other ML-enabled systems.  

 \end{abstract}

 \section{Introduction} \label{Introduction}

Driving automation systems (DAS), including both autonomous driving (AD) and advanced driver assistance systems (ADAS), are software systems designed to augment or automate aspects of vehicle control~\cite{mallozzi2019autonomous}. DAS have long been a domain of interest. However, the increased capabilities and usability of machine learning (ML) have subsequently improved the capabilities of---and interest in---such systems. Research advances have produced improved comfort and safety, and reduced fuel and energy consumption, emissions, and travel time~\cite{mallozzi2019autonomous}.

\begin{figure} [!t]
    \centering
     \includegraphics[bb=00 00 870 270, width=0.7\linewidth]{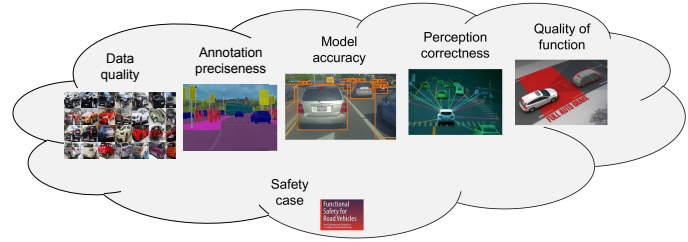}
    \caption{Conceptual model of quality transitions from data collection to the quality of the automotive function.}
    \label{fig:ConceptualModelQualityTransitions} \vspace{-20pt}
\end{figure}

DAS functionality depends on the correctness and the integrity of perception systems that blend ML-based models and traditional signal processing\footnote{In this paper, we focus specifically on ML-based perception systems for DAS, but often use the term \textit{perception systems} as shorthand.}. The usage of ML for perception relies on a large quantity of data. Data quality, context, and attributes---as well as annotation quality---have a significant impact on the resulting system quality. However, it is difficult to make direct connections between data, annotation, ML model quality and the resulting functional quality of a perception system (e.g., between the boxes in Fig.~\ref{fig:ConceptualModelQualityTransitions}). The inherent uncertainty of ML---coupled with the desired levels of data quality and coverage---creates substantial process and requirements engineering (RE) challenges in perception system development~\cite{borg2018safely}.

RE is an important foundational element of quality assurance and safety engineering. RE plays a critical role in perception system development by enabling explicit capture of safety and quality requirements, supporting communication, recording functional expectations, and ensuring that standards are followed. Recent research has explored RE challenges for ML systems, e.g.,~\cite{belani2019requirements,vogelsang2019requirements}. However, such challenges have not been thoroughly explored in the context of perception systems for DAS. Addressing this gap is necessary to advance practices in both this domain and in the broader context of RE for AI. 

To explore important topics and challenges for perception systems, we have conducted an interview study with 19 expert interviewees from five companies working in various DAS roles. We analyzed interview data using thematic coding to produce eight major themes: perception, requirements engineering, systems and software engineering, AI and ML models, annotation, data, ecosystem and business, and quality. Here, we analyze data collected as part of the RE theme, and explore critical RE topics and challenges for perception systems\footnote{A recent submission has used the same study data, but focuses on the annotation, data, and ecosystems and business themes~\cite{heyn2022automotiveperception}.}.

Our findings indicate that practitioners have difficulty breaking down specifications for the ML components. In practice, individuals report that they use scenarios, operational design domains (ODDs), and simulations as part of RE. Practitioners experience RE challenges related to uncertainty, ODD detection, realistic scenarios, edge case specification, traceability, creating specifications for data and annotations, and quantifying quality requirements.

By summarizing the views and challenges of different experts on RE for ML-enabled perception systems, our results are valuable for practitioners working to advance this area. Additionally, our findings contribute to improving RE knowledge more broadly in other domains reliant on ML.
 \vspace{-0.2cm}
\section{Related Work} \label{Related Work}

\noindent\textbf{RE for ML:} Recent research has focused on how RE could or must change in the face of rising use of ML. Systematic mapping studies on RE for ML identified new contributions in this area, including approaches, checklists, guidelines, quality models, classifications and evaluations of quality models, taxonomies, and quality requirements~\cite{villamizar2021requirements,ali2022systematic,habibullah2022non}. Ahmad et al. investigated current approaches for writing requirements for AI/ML systems, identified tools and techniques to model requirements for AI/ML, and pointed out existing challenges and limitations in this area~\cite{ahmad2021s}.
Belani et al. identified and discussed RE challenges for ML and AI-based systems, and  reported that identifying NFRs throughout the software lifecycle is one of the main challenges~\cite{belani2019requirements}. Heyn et al. used three use cases of distributed deep learning to describe AI system engineering challenges related to RE~\cite{heyn2021requirement}, including context, defining data quality attributes, human factors, testing, monitoring and reporting.

\noindent\textbf{RE for Vehicles and DAS:}
Significant research has been performed on RE for vehicles. Liebel et al. identified challenges in automotive RE with respect to communication and organization structure~\cite{liebel2018organisation}. 
Pernstal et al. stated that RE is one of the areas most in need of improvement at automotive original equipment manufacturers (OEMs), and also identified the ability to communicate via requirements as important~\cite{pernstaal2013software}. Allmann et al. also noted requirements communication as a major challenge for OEMs and their suppliers~\cite{allmann2006requirements}.
Mahally et al. identified that requirements are the main enablers and barriers of moving towards Agile for automotive OEMs~\cite{m2015barriers}.

Research has also looked specifically at RE for AD, e.g., providing an overview of AD RE  techniques~\cite{staron2019requirements}, 
Riberio et al. identified AD RE challenges addressed by the literature, and identified the languages and description styles used to describe AD requirements, with special attention given to NFRs~\cite{ribeiro2022requirements}. 
Heyn et al. investigated challenges with context and ODD definition in ML-enabled perception systems~\cite{heyn2022setting}, including a lack of standardisation for context definitions, ambiguities in deriving ODDs, missing documentation, and lack of involvement of function developers while defining the context. \r{A}gren et al. identified six aspects of RE that impact automotive development speed, moving toward AD~\cite{aagren2019impact}.

 \vspace{-0.2cm}
\section{Methodology} \label{Methodology}

\begin{figure}[h!]
    \vspace{-0.5cm}
    \centering
      \includegraphics[bb=00 00 580 110, width=\linewidth]{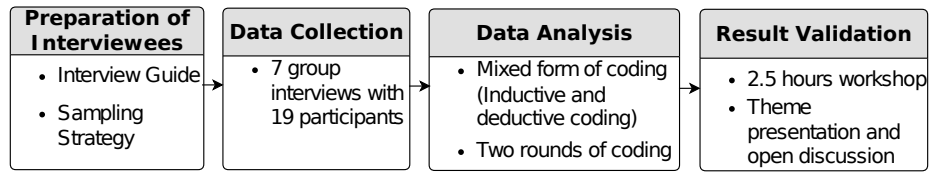}
      \vspace{-25pt}
    \caption{Overview of interview study.}
    \label{fig:study_overview} \vspace{-20pt}
\end{figure}

\noindent Our study is guided by the following research questions:
\begin{itemize}
    \setlength{\itemsep}{1pt}
  \setlength{\parskip}{0pt}
  \setlength{\parsep}{0pt}
    \item [\textbf{RQ1:}] What are the RE-related topics of interest for perception systems for DAS?
    \item [\textbf{RQ2:}] What challenges are experienced in RE for  perception systems for DAS?
\end{itemize}
To address these questions, we conducted seven group interviews with 19 expert participants from five companies that are currently working in ML-based perception systems for DAS. Fig.~\ref{fig:study_overview} gives an overview of the interview study.









\noindent\textbf{Data Collection:}
We used semi-structured group interviews with a set of predetermined open-ended questions\footnote{The interview guide can be found at: \url{https://doi.org/10.7910/DVN/HCMVL1}.} to keep enough freedom to add follow-up with additional questions. 
The interviews were conducted between December 2021 and April 2022 via Microsoft Teams, and lasted between 1 hour and 30 minutes to 2 hours. We recorded all interview sessions with the permission of all participants; then transcribed, and anonymized the recordings for analysis. At least three researchers were present in each interview, with the same two researchers in all interviews to maintain consistency.

\begin{table}[h!]
\vspace{-1cm}
\centering
\scriptsize
\caption{Overview of conducted interviews (same as~\cite{heyn2022automotiveperception})}
\label{tab:participants}
\begin{tabular}{cll}
\hline
\multicolumn{1}{c}{\textbf{\begin{tabular}[c]{@{}c@{}}Interview\end{tabular}}} & \multicolumn{1}{l}{\textbf{Field of work}} & \multicolumn{1}{l}{\textbf{Participants}} \\ \hline
\rowcolor[HTML]{EFEFEF} 
A & Object detection & Product owner \\
B & Autonomous Driving & \begin{tabular}[l]{@{}l@{}}Product owner, test engineer, ML engineer, \\ software developer \end{tabular} \\
\rowcolor[HTML]{EFEFEF} 
C & Vision systems & \begin{tabular}[l]{@{}l@{}}System architect, product owner, \\ requirement engineer, deep learning engineer \end{tabular} \\ 
D & AD and ADAS & \begin{tabular}[l]{@{}l@{}}System engineer, manager AD \end{tabular} \\
\rowcolor[HTML]{EFEFEF} 
E & Testing and validation AD &\begin{tabular}[l]{@{}l@{}}System architect, two product owners, \\ compliance officer, data scientist \end{tabular} \\
F & Data annotations &\begin{tabular}[l]{@{}l@{}} AI engineer, data scientist \end{tabular} \\
\rowcolor[HTML]{EFEFEF} 
G & Autonomous Driving & \begin{tabular}[l]{@{}l@{}} System safety engineer \end{tabular} \\ \hline
\end{tabular} \vspace{-15pt}
\end{table}

A summary of the participants is shown in Table~\ref{tab:participants}. We chose participants who posses experience with ML, perception systems for DAS, software and systems engineering, RE, or data science, or who were working in the DAS industry. The sampling method was a mix of purposive, convenience, and snowball sampling. We sent open calls to the Swedish automotive industry, and our known contacts, then we asked the interviewees for further contacts. Our participants work with different aspects of DAS. 

We started by asking for demographic information about the participants. 
We then showed them Fig.~\ref{fig:ConceptualModelQualityTransitions}, asking for their feedback and using the figure to ground further discussions about how functional requirements relate to requirements on data and data annotation. We asked further questions about their requirements documentation, safety issues, and quality. Although we carefully chose interview participants, the opinions of the individual interviewees do not necessarily reflect the overall opinion of their companies. 
Due to the sensitive nature of information provided by interview participants and their respective companies, we are unable to disclose the raw interview data or specific details about ways of working.
Finally, in a 2.5-hour workshop with roughly 20 participants, many of whom were interviewees, we presented and discussed our findings with illustrative quotes.


\noindent\textbf{Data Analysis:}
We applied thematic analysis, as per Saladana~\cite{saldana2021coding}. We used a mixed form of coding, where we started with a number of high-level deductive codes based on the interview questions, then we started inductive coding, adding new codes while going through the transcripts. At least three of the researchers worked together to code each of the transcribed interviews. We observed saturation after five interviews, as not many new inductive codes emerged. In a second round of coding, a new group of at least two researchers reviewed the interview transcripts and verified the codes. Finally, we used pattern coding to identify emerging themes and sub-categories. 
To illustrate our points, we use a number of interview quotes.  For increased anonymity, participants are assigned a random identifier, such that P1 does not necessarily match to interview A. In this paper, we focus on findings specifically in the RE theme.  Heyn et al. have analyzed the ecosystem and business, data, and annotation themes~\cite{heyn2022automotiveperception}. Further themes will be analyzed in future work.



 \vspace{-0.3cm}
\section{Results} \label{Results}



\begin{figure}[h!]
\vspace{-0.5cm}
    \centering

    \includegraphics[bb=00 00 620 220, width=\linewidth]{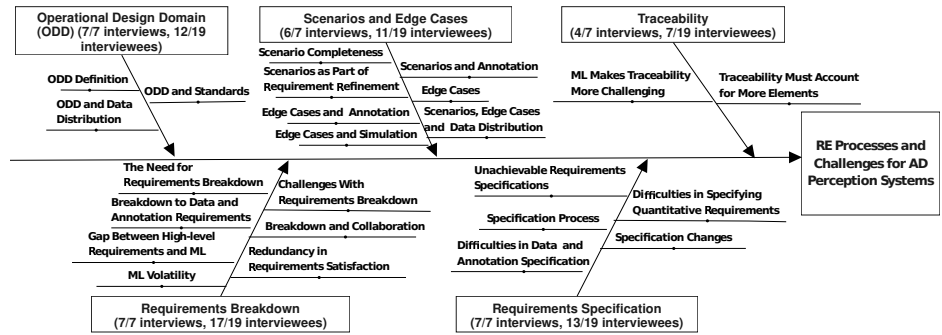}
    \caption{Mind map illustrating relevant RE topics and challenges for DAS perception systems.}
    \label{fig:CauseEffectRequirementsEngineeringTheme} \vspace{-30pt}
\end{figure}

Based on the thematic analysis, we divide the RE theme into sub-themes---``Operational Design Domain (ODD), ``Scenarios and Edge Cases'', ``Requirements Breakdown'', ``Traceability'' and ``Requirements Specification''---and important topics within each sub-theme. The sub-themes and topics are summarized in Fig.~\ref{fig:CauseEffectRequirementsEngineeringTheme}. Our themes reflect both RE topics and challenges, addressing both RQ1 and RQ2.
We also note how many interviewees discussed the sub-theme. 
These sub-themes and topics answer RQ1, identifying relevant RE-related topics in perception systems.  We use these results to identify which topics are, or contain, specific challenges (RQ2) in Sec.~\ref{Discussion}.

\vspace{-0.2cm}
\subsection{\textbf{Operational Design Domain (ODD)}}

An ODD is a description of a domain that a DAS will operate in---e.g., the road or weather conditions. As part of RE, one needs to define not only requirements, but assumptions about the domain, context, and scope of operation. Operational context and scope for perception systems is particularly important as the intensity of hazards depends upon the current ODD. ODD-related topics came up in all interviews and were discussed by 63\% of the participants.  



\noindent\textbf{ODD Definition:} ODDs should be captured as part of the requirements specification. Several interviewees mentioned ODD detection---where the system detects that a certain ODD is currently applicable for a DAS function---and ODD exit detection---when the ODD is no longer applicable. ODD detection requires information on what to detect and detection accuracy. For example, on highways, DAS needs to detect different dynamic objects than in urban areas. 

\noindent\textbf{ODD and Standards:} Interviewees state that ODDs are critical, and therefore, it is desirable to follow a standard or process for specifying and defining ODDs. 
This need has been recognized and new initiatives for the definition of ODD exist, e.g., the interviewees mention the PAS-1883 standard, and we are aware of other standards (e.g., ISO 21448/SOTIF) that include ODDs.

\noindent\textbf{ODD and Data Distribution:} One interviewee stated that data distribution requirements are highly influenced by ODDs. 
For example, camera data can be classified according to descriptions in the ODD, and this mapping can reveal missing data, driving further data collection. 
As it is not feasible to collect data in all possible contexts, it is necessary to have an efficient sampling process covering the most common ODDs. 
\vspace{-10pt}
\interviewquote{If the performance of the model is not good enough in some part of the ODD, for instance during the night or snow weather and so on, then we can select more samples from those areas.}{P16}
\vspace{-10pt}


Another interviewee pointed out that although ODDs drive data collection, collecting certain types of data required by the ODD can still be very difficult.  
\vspace{-10pt}
\interviewquote{
... mining for specific use cases. For instance, it is not easy to collect data that contains animals in it. You need some way to mine and find those specific frames which will be sent for annotations and then be used during training.}{P16}
\vspace{-10pt}

 \vspace{-0.2cm}
\subsection{\textbf{Scenarios and Edge Cases}} \label{Scenarios and Edge Cases}

Several interviewees described how scenarios are crucial as part of the requirements specification process. In this context, scenarios describe specific operational paths and conditions for a vehicle, and one ODD may include a number of scenarios. 
As such, although there are links to scenario-based requirements methods\cite{sutcliffe2003scenario}, there are also clear differences. Scenarios and edge cases came up in 86\% of interviews and were discussed by 58\% of participants. 



\noindent\textbf{Scenario Completeness:} 
It is important that perception systems perform correctly and that the vehicle handles failures in as many scenarios as possible. As such, scenarios can help in requirements derivation.
\vspace{-10pt}
\interviewquote{If we refer to the classic system engineering process, I think nowadays it's quite hard ... we are trying to use the scenario to derive the requirements. If we
... see the features or the distribution of the scenarios based on the data from the real world. Then we can derive the high-level requirements based on that data, the scenario database.}{P4}
\vspace{-10pt}


\noindent One interviewee stressed the difficulty of defining and assessing coverage. 
\vspace{-10pt}
\interviewquote{How do you define coverage? ... What is the scenario space for pedestrian children? Is it based on how the area you have annotated looks inside of your bounding box? Do you parameterize it on the size of the bounding box, parameterized on conditions around you? How would you divide that space and define it in a way that allows even measures? Have I covered not just enough children, but also enough variety of children? }{P18}
\vspace{-10pt}

\noindent\textbf{Scenarios and Annotation:} 
Even if all important scenarios are reflected in training data, annotation errors may result in unsafe behavior---e.g., a perception system may recognize a human as a tree during a snowy or rainy day. 
\vspace{-10pt}
\interviewquote{We'll pick out some scenarios that we feel (are) likely not correct, for instance, if it's a rainy night, then maybe the annotator is not annotating (people) as accurately as in the day.}{P8}
\vspace{-10pt}


\noindent\textbf{Scenarios as Part of Requirement Refinement:} Our results show that testing through scenarios enables iterative requirements refinement. Engineers iteratively refine their expectations of correct behavior by examining scenarios and capturing observations from simulation or in the field.
\vspace{-10pt}
\interviewquote{... we have to learn through testing, so probably it will start with some rough set of requirements, some obvious setting requirements. Then we will, through real-world testing, discover and learn exactly how we want to behave.
}{P2}
\vspace{-10pt}
\vspace{-10pt}
\interviewquote{It seems like a test-driven development process ...  we have the scenarios to drive the development and give more input and also we get the benefit of testing.}{P4}
\vspace{-10pt}

\noindent\textbf{Edge Cases:} 
Interviewees stated that, in addition to normal scenarios, it is crucial and challenging to deal with edge cases. The interviewees used subtly different terms, such as edge cases, rare cases, and cases that occurred very infrequently. We use the term ``edge cases'' for simplicity. These cases may be missed by studying data distributions, but are very critical to ensure safety. 

\vspace{-10pt}
\interviewquote{The cars ... will end up in situations that no one could predict, that we've never seen before, and somehow we need, even in this situation, one individual car needs to perform better than a human driver, and human drivers are real good at handling edge cases. 
The neural networks will not do that.}{P13}

\vspace{-10pt}
\noindent\textbf{Edge Cases and Annotation:} Edge cases cause issues by creating confusion among annotators. Data from edge cases is often annotated inconsistently. The topic of annotation is explored in more detail by Heyn et al.~\cite{heyn2022automotiveperception}. 
\vspace{-10pt}
\interviewquote{We label whether a vehicle is in our lane or not. But how should you? You can think of so many corner cases when you are out driving. When you are doing a lane change. Which lane are you in then, and how would you then place all the other vehicles or lane lines? Maybe there are double lane lines and which is valid and which is not?  This leads to a lot of confusion among annotators.}{P17}
\vspace{-10pt}


\noindent\textbf{Scenarios, Edge Cases and Data Distribution:}  One interviewee pointed out that scenarios, and especially rarer edge cases, are important for driving data collection efforts as part of having an effective data distribution. How well edge cases are covered can be an important development metric.


\noindent\textbf{Edge Cases and Simulation:} Interviewees stated that collecting data points for particular scenarios from the real world is necessary, but is particularly difficult for edge cases. This makes simulation challenging, as for safety-critical edge cases, practitioners have difficulty safely gathering enough data to run realistic simulations. This makes the process of iterative requirements refinement, as described previously, difficult for requirements associated with edge cases.

\vspace{-0.2cm}
\subsection{\textbf{Requirements Breakdown}} \label{Requirements Breakdown}

Requirements breakdown can involve both refining or decomposing requirements. Requirements breakdown was brought up as a topic in all interviews and was discussed by 90\% of participants. 

\noindent\textbf{The Need for Requirements Breakdown:}
We see evidence that a traditional requirements breakdown is followed for perception systems. At least one participant spoke of splitting the problem to reduce complexity.  
\vspace{-10pt}
\interviewquote{We need to split the problem. We can't do all work at the same time on the complete problem.}{P12}
\vspace{-10pt}






\noindent Another participant described an architectural-oriented breakdown.
\vspace{-10pt}
\interviewquote{Let us say you don't want to collide with an object more than once in a billion hours. This is your top requirement and then you need some kind of architecture or idea of what your system looks like. That should realize this safety goal. This is where we typically come up with a functional architecture, and we start to break down the requirements of the parts of that functional architecture. Then we work. We refine it. The functional architecture becomes a system or logical architecture and we break it down into smaller and smaller pieces.}{P7}
\vspace{-10pt}

\noindent Others describe the importance of separation of high-level requirements from technical requirements to have an upper layer that is resilient to change.  
\vspace{-10pt}
\interviewquote{To me, at least the function level will be the same in 100 years because there's no need that you change it. If your function doesn't change, because today you satisfy that function by combustion engine, in the next 50 years by electric, and in the next, I don't know, 100 years by something more intelligent ... 
By changing your technical system level specifications, you still can satisfy your function.
}{P19}
\vspace{-10pt}


\noindent\textbf{Challenges with Requirements Breakdown:} 
Participants commented on the challenges of connecting high-level requirements to low-level requirements and general challenges with requirements breakdown in this context. 
\vspace{-10pt}
\interviewquote{I would say we're working with that challenge and, not that it's an easy one, but we do believe that it's necessary to connect the top-level requirements or the quality of the function, and to map that to quantitative or performance requirements on, for example, perception, precision, and control.}{P13}
\vspace{-10pt}
\vspace{-10pt}
\interviewquote{What you can do is interact the most closely with ... some component, maybe in perception, and these are the ones who would place direct requirements on the previous component, so it is to me a bit of a hierarchical model to approach the difficulties in breaking down the final safety goal to the early stages in our processing chain. I think one tricky thing is, that it's a hierarchical way in some ways, but you also have to go in both directions in that hierarchical model.
}{P6}
\vspace{-10pt}


Several interviewees report that traditional requirements breakdowns cannot be easily applied.  
\vspace{-10pt}
\interviewquote{For sure, we will not start with the classical software approach, where you start with some requirements and then keep breaking those down and through the V-Model because it will be impossible to capture the behavior of autonomous vehicle with requirements.} {P2}
\vspace{-10pt}

\noindent\textbf{Breakdown to Data and Annotation Requirements:} Interviewees explained that, although linking functional requirements to system accuracy is often possible, breaking functional requirements into data and annotation requirements is more difficult.  
\vspace{-10pt}
\interviewquote{
Working with system level requirements, I can look at function requirements and figure out roughly what kind of accuracy we need ...
That does not necessarily mean that I can tell how precisely annotation has to be, 
because I need to know how the software works to figure that out. 
Another translation needs to happen where I gave my requirements to the developers and they have to figure out what kind of accuracy they need from the data to 
meet the system requirements and with so many translations on the way, it is easy for things to get lost somewhere.}{P6}
\vspace{-10pt}
\vspace{-10pt}
\interviewquote{...it is difficult to write good requirements on data quality and annotation preciseness and have those links all the way up to feature requirements (Fig.~\ref{fig:ConceptualModelQualityTransitions}). 
Which I think 
is because of the dimensionality of the problem. 
The input space is so enormous that it's really tricky to get a single set of requirements there.}{P15}
\vspace{-10pt}

\noindent\textbf{Breakdown and Collaboration:} Challenges arise when teams collaborate to specify quality requirements. 
\vspace{-10pt}
\interviewquote{
Creating one function would involve multi-team collaboration usually. I guess it's not as easy as evaluating your own system when other people are kind of involved, so you have to come up with scenarios and things to test your algorithms with and could try to come up with a plan. 
}{P4}
\vspace{-10pt}

Frequent and direct interaction with the stakeholders can reduce this difficulty and help engineers to identify the requirements. In this case, stakeholders have internal roles in the perception system development.  
\vspace{-10pt}
\interviewquote{I think it is a lot of interaction with direct stakeholders in the end ... because the direct consumers of whatever you are producing know exactly what they need to fulfill their own requirements from their own stakeholders. So the negotiation across these interfaces is where the most interaction happens. 
}{P9}
\vspace{-10pt}

\noindent\textbf{Gap Between High-level Requirements and ML:}  
When breaking down high-level requirements to very specific requirements on the ML-based perception system, results show that traditional RE practices are able to be applied up to a certain point - even though challenging. However, the breakdown for the ML based components is particularly challenging.
As such, there are boundaries within the system where requirements methods change.  
\vspace{-10pt}
\interviewquote{If we talked about some other requirements or specifications not for the AD stack. 
... those things still can follow the traditional way 
for critical system. 
... if we distinguish those two parts, ... 
for the black box or part or AD business part, it's hard to follow, but for the rest we still can leverage the classic knowledge.}{P4}
\vspace{-10pt}

We see that it is difficult to specify requirements for the whole perception system. However, there are often still requirements---in terms of various performance metrics---at a high-level. 
\vspace{-10pt}
\interviewquote{
If we say the requirements were specified for the entire AD stack, I think it's quite hard to have very precise or detailed specifications for all functions, but actually, we have some high-level metrics like safety, performance, functionality, or traffic comfort metrics ... 
We have something, but they are very different from the traditional understanding of the specification.}{P4}
\vspace{-10pt}




\noindent\textbf{Redundancy in Requirements Satisfaction:} One interviewee described how requirements are allocated to ensure redundancy in the solution. 
\vspace{-10pt}
\interviewquote{We typically try to break down the problem to come up with redundant solutions. You would have one algorithm using one sensor, which has some capacity to detect the pedestrian, and then use another algorithm and another algorithm in parallel. And you use another sensor and ...  
decompose the problem such that ... 
it's very unlikely that all of them would miss this pedestrian. That's a way to try and get reasonable requirements on every perception component.}{P6}
\vspace{-10pt}

\noindent\textbf{ML Volatility:} 
One interview pointed out, due to dependencies between components and the volatile nature of ML, changes in the ML model can cause drastic changes in other parts of the system.
\vspace{-10pt}
\interviewquote{
People sometimes start setting requirements on sensors, and then start setting requirements on data, and calibration accuracy, 
and then also 
on annotation, preciseness, and that somehow should influence the model accuracy. 
Maybe one problem we have with ML is that, if there are things slightly off, it cannot just lead to a slight degradation, but to complete degradation of the entire system.}{P17}
\vspace{-10pt}

\vspace{-0.2cm}
\subsection{\textbf{Traceability}} \label{Traceability}

37\% of interviewees, in 57\% of the interviews, brought up points related to traceability in perception systems.

\noindent\textbf{ML Makes Traceability More Challenging:}
Known traceability challenges are exacerbated by the use of ML and associated data. 
Interviewees described that when systems or modules fail to meet particular key performance indicators (KPIs), tracing the source of the issue is difficult due to the combination of ML models and traditional code. Traceability was discussed in four out of our seven interviews and by seven out of 19 participants. 

\vspace{-10pt}
\interviewquote{I think what is important at the end is 
the KPIs on the rightmost features of the figure (Fig.~\ref{fig:ConceptualModelQualityTransitions}). Then if you want to track down why it is not working, 
it's not very easy to find which module is not working as supposed to, or maybe it works, but 
in a combination of something else, 
it creates some kind of strange behavior. 
}{P14}
\vspace{-10pt}

\noindent\textbf{Traceability Must Account for More Elements:}
It is important 
that traceability be maintained not just between code and requirements, but also with ML elements---e.g., models and datasets---that determine the overall functionality.

\vspace{-10pt}
\interviewquote{
I think it is important to keep track of exactly which data was used to train the model, 
and be able to also show that 
to the general public if needed, right? ... having traceability all the way through development 
 is something we aim for.}{P8}
\vspace{-10pt}




Typically, trace links would link to typical elements like requirements and safety goals, but now they should also link to scenarios.
\vspace{-10pt}
\interviewquote{I don't want to say something that is wrong, you need this traceability, and then when you trace back you see that, OK, I had a safety goal that was talking about this specific scenario. 
}{P19}
\vspace{-10pt}

\vspace{-0.2cm}
\subsection{\textbf{Requirements Specification }} \label{Requirements Specification}

Aspects of documentation and requirements specification were discussed in all interviews, and by 68\% of participants. 






\noindent\textbf{Unachievable Requirements Specifications:} Two interviewees mentioned that sometimes clients provide unachievable requirements, even though requirements specifications are clear and precise.
\vspace{-10pt}
\interviewquote{Sometimes clients come to us with a very well written set of requirements, 
like we want this annotator and want this precision or accuracy ... 
Then they send us data. But when we start looking at the data, it turns out that, given this data, these requirements are basically impossible to meet.}{P18}
\vspace{-10pt}


\noindent\textbf{Difficulties in Specifying Quantitative Requirements:} Due to confidentiality, interviewees were not able to elaborate on specific target levels for quantitative requirements. However, they did reflect generally about the difficulty in determining quantitative quality targets.  
\vspace{-10pt}
\interviewquote{... for model accuracy, what does success look like in 
functional safety? If you can recognize 99\% rebounding boxes of possessions, is it good enough? If you have a recall of 100\%, but your precision is only 50\%, would that be good enough?}{P17} 
\vspace{-10pt}

\noindent\textbf{Specification Process:}
One interviewee emphasized that documentation of the rationale and goals of the project can serve as a form of requirement specification.

\vspace{-10pt}
\interviewquote{I think it's valuable to actually document after what principles you're working, document the problem you're trying to solve and that is basically a set of requirements, even if they're not necessarily traceable upwards all the way.}{P15}
\vspace{-10pt}



\noindent\textbf{Specification Changes:} The uncertain and highly iterative nature of perception systems and their development environment means that specifications are particularly prone to change. 
\vspace{-10pt}
\interviewquote{Requirements at any level are not something that is static. They should reflect your current best interpretation. 
These things can change because your understanding 
or your development process changes or the environment changes because there are suddenly new demands on how something is supposed to perform or you learn something new about the system or its environment. 
}{P15}
\vspace{-10pt}

\noindent\textbf{Difficulties in Data and Annotation Specification:} One interviewee said that specifying data requirements is difficult and different from functional specification, as it is hard to identify features and ensure data quality upfront. 
\vspace{-10pt}
\interviewquote{It's very different how you write a data specification ... it's hard to know what the future expects and what type of classes we want and how we want to combine certain objects ... we future proof our datasets quite well by specifying. We do specify a lot of classes.}{P5}
\vspace{-10pt}

Another interviewee reported that it is difficult to specify quality (non-functional) requirements on data and annotation, and to understand how qualities affect model performance.
\vspace{-10pt}
\interviewquote{
I work a lot with image quality 
before any ML is involved. Even that is 
very difficult to quantify. We can have very much right objectively measurable requirements on image quality, sharpness. 
Then how those translate to the actual performance of a ML algorithm is not at all linear.}{P16}
\vspace{-10pt}

Another participant described challenges in specifying requirements for data annotation when dealing with external partners. It is difficult to have an upfront, detailed specification of data classes and accuracy levels. Instead, data specification needs to be developed iteratively and experimentally with suppliers.  



 \vspace{-0.2cm}
\section{Summary and Discussion} \label{Discussion}
\vspace{-0.2cm}


\noindent\textbf{RE Topics (RQ1):} We have identified a number of RE topics in Sec.~\ref{Results}, as summarized by Fig.~\ref{fig:CauseEffectRequirementsEngineeringTheme}. These topics can be seen as a sort of check-list when working with ML-based perception systems---a list of issues that should be considered.

Our interviewees emphasize that the definition and limits of ODDs are an integral part of perception systems, and these ODDs have important impacts on data requirements and collection, confirming findings in Heyn et al.~\cite{heyn2022setting}. Similarly, perception systems development relies heavily on the use of scenarios and associated edge cases. Such scenarios play a key role in dictating annotation, data collection and simulation. As part of the RE process for perception, it is particularly important to capture edge case scenarios, and these edge cases also play an important role in annotation, simulation, and data collection.

\noindent\textbf{RE for Perception System Challenges (RQ2):} 
Our results indicate that \textbf{ODD detection and ODD exit detection} are challenging, as this requires information not only about what to detect in the environment, but also how to detect it and the accuracy of the detection. In addition, \textbf{data requirements} are highly influenced by the content of an ODD, therefore ODDs can be used to evaluate whether a data distribution is sufficient for good ML model performance. However, it is not always easy to \textbf{collect the data} specified by ODDs. Heyn et al. also emphasized the importance of ODDs in DAS, and noted 
the lack of a common definition for ODDs~\cite{heyn2021requirement}.  Our participants go further and mention the need for ODD standardization (and efforts in that regard).


    One major challenge is that simulations should reflect \textbf{realistic scenarios}, echoed by Acuna et al.~\cite{acuna2021towards}. For ensuring safe perception, the collected data and scenarios must be thorough, and the perception system must avoid failure in all scenarios. In addition to covering normal scenarios, it is important to \textbf{specify edge cases} among scenarios, which are then used to determine data distributions. However, edge cases introduce challenges as they create \textbf{confusion among annotators} and are challenging to \textbf{test in reality} due to safety concerns.

\textbf{Breaking down requirements} for data and annotations  can be very difficult, and additional challenges are introduced due to requirements dependencies and the need for multiple teams to collaborate.  In general, we believe that the \textbf{gap between standard RE methods and ML components} is both a technical gap and a gap in training and backgrounds, as the ML components are often engineered by data scientists without a software background.

Difficulties in breakdown, ML opaqueness, as well as the the introduction of more elements to trace (e.g., ODDs, scenarios, training data), make it difficult to establish \textbf{traceability}. These challenges are in addition to the known challenges with motivating and using traceability in practice~\cite{wohlrab2016collaborative}. 

Creating \textbf{specifications for data and annotations} is challenging, as it is difficult to have an upfront specification for data classes, e.g, pedestrians and crosswalks. Furthermore, sometimes ML components are assigned unrealistic and \textbf{unachievable requirements}. Although requirements change is a frequently acknowledged RE problem~\cite{jayatilleke2018systematic}, with perception systems, the \textbf{level of uncertainty and change} is particularly high due to uncertainty about the system, including ML, and the environmental targets. \textbf{Quantifying quality requirements} (e.g., accuracy) is also particularly challenging in perception systems, echoing the results of Vogelsang and Borg~\cite{vogelsang2019requirements}. 


Some of these challenges are relatively new from an RE perspective (e.g., ODD detection, missing edge case), while others have been long recognized (e.g., traceability~\cite{wohlrab2016collaborative}, specification changes~\cite{jayatilleke2018systematic}).
As mentioned, three additional themes from the same study are reported and analyzed by Heyn et al.~\cite{heyn2022automotiveperception}. Although the article focuses on different themes, the qualitative topics covered in that work and our work here have some overlap, particularly in topics related to data and annotation.  However, here, the topics of data and annotation are approached from an RE perspective, while the other article takes an ecosystems and process view on topics and challenges related to perception systems in DAS.

Although the focus of this work has been on perception systems, we believe that many of the RE practices and challenges found would apply more generally to other domains reliant on ML. For example, challenges breaking down specification would hold due to the volatility and opaqueness of ML. Further work should contrast RE challenges and practices in other ML-enabled domains. 
 \vspace{-0.2cm}
\subsection{Threats to Validity} \label{Threats to Validity}
\vspace{-0.1cm}

\noindent\textbf{Internal Validity:}
We internally peer-reviewed the interview guide and conducted a pilot interview to improve the guide and process. We sent a preparation email to all the interview participants with the details and purpose of the interview study. To maintain consistency in the interview process, at least three authors conducted each interview, with two authors present in all interviews. 

All interviews were conducted in English, and the auto-generated transcripts were `fixed' by authors by listening to audio recordings and correcting any transcription errors. Note that the working language of each company was English, so the language should not have created barriers.  

Although qualitative coding always comes with some bias, we mitigated this threat by following established literature~\cite{saldana2021coding}, coding in multiple rounds, using inductive and deductive codes, and having multiple authors participate in each round of coding, with in-depth discussion on code meanings and assignments.   

\noindent\textbf{External Validity:} 
We used a  mixture of purposive and snowball sampling. As our study needed a certain set of expertise to answer our questions, we could not conduct random sampling, using our networks and their contacts.  Still, due to the size of the study, with participants covering a wide variety of roles with varying experience levels, covering differing company roles and sizes in the perception system ecosystem, we believe we have a relatively representative sample.  
Furthermore, we argue that we reached a sufficient point of saturation with our interview data, as we noticed a sharp decline in emerging codes after analyzing the fifth group interview. 

Note that one cannot link participants to interviews and companies, this is done deliberately to protect the anonymity of our participants.  Although this may affect transferability of our results, we feel this level of anonymity does not greatly hurt our results.  
Though our study results are limited to perception systems in DAS, we  argue that some findings can apply to other safety-critical or perceptions systems.  This applicability should be explored in future studies.  

 \vspace{-0.4cm}
\section{Conclusion} \label{Conclusion}
\vspace{-0.2cm}

Our study investigated RE practices and challenges during the development of PS. We interviewed 19 participants from five companies and identified a number of RE practices and challenges that impact heavily the functional safety assurance of PS for DAS. The results of this study suggest future research directions in RE and ML to mitigate the challenges practitioners are facing.
\vspace{-0.3cm}

\subsection*{Acknowledgements}
\vspace{-0.2cm}
Support for this project was provided by Vinnova pre-study 2021-02572.  We thank all participants.
\vspace{-0.3cm}

\bibliographystyle{splncs04}

\end{document}